\documentstyle{article}
\newcommand{\bra}[1]{\left \langle #1 \right \vert }
\newcommand{\ket}[1]{\left \vert #1 \right \rangle}

\begin{document}
\title{Delocalized Qubits as a Computational Basis in the System of Interacting Spins}
\author{Alexander R. Kessel and Vladimir L. Ermakov \thanks{e-mail: ermakov@dionis.kfti.knc.ru}\\
{\it Kazan Physico-Technical Institute, }\\
{\it Kazan Science Center, }\\
{\it Russian Academy of Sciences,} \\
{\it Kazan 420029 Russia}}
\date{\today}
\maketitle
\begin{abstract}
It is suggested to map the qubits into solid state NMR spin system collective states 
instead of the states of the individual spin. Such an approach introduces the stable 
computational basis without any additional actions and allows to obtain the universal 
set of quantum gates, which operation time is determined only by a RF pulse duration. 
\end{abstract}

PACS No.: 03.65.-w, 03.67.-a, 76.60.-k.


\section{ Introduction}

\par Nuclear magnetic resonance (NMR) provides an excellent proving ground for 
testing different quantum information processing (QIP) ideas. Due to combination 
of its good developed theory and sophisticated experimental technique the 
realization of the simplest quantum algorithms using standard NMR spectrometers 
turned out to be possible. At present time the achievements of liquid state 
(LS) NMR QIP are far beyond the capabilities of any other experimental methods. 

\par This success of LS NMR QIP is due also to the fact that a clear correspondence 
was found between the abstract notion of a quantum mechanical bit - qubit and real 
physical object - nuclear spin. Namely, two qubit states $\ket{0}$ and $\ket{1}$ are 
mapped onto two possible spin 1/2 projections. As a computational basis the eigenfunctions 
$\ket{m}$ of the Z-components of a nuclear spin are used, which commutates with the full 
Hamiltonian. For example, for two spin system this computational basis is written 
as follows:

\begin{equation} \label{two_spin_system_computational_basis}
\begin{array}{cc}  
\ket{00}=\ket{m_{1}= -1/2, m_{2}= -1/2},&\ket{01} = \ket{m_{1}= -1/2, m_{2}= +1/2},\\
\ket{10}=\ket{m_{1}= +1/2, m_{2}= -1/2},&\ket{11} = \ket{m_{1}= +1/2, m_{2}= +1/2}. 
\end{array}
\end{equation}

\noindent where $\ket{m_{1}}$ and $\ket{m_{2}}$ are the nuclear spin Z-components 
eigenfunctions. 

\par However, already now one can see the LS NMR QIP limitations and the next step is 
going to be the solid state (SS) NMR QIP \cite{Jones}-\cite{Cory}. 
The SS NMR essential feature is the existence 
of the spin-spin (exchange and dipole-dipole) interaction terms in the total Hamiltonian,
which are not averaged to zero by thermal motions, and which contain not only Z, but 
also X and Y spin components. Due to this fact a single spin orientation becomes bad 
integral of motion. This means that such system stationary states do not correspond 
to definite values of the spins Z-component and the functions 
(\ref{two_spin_system_computational_basis}) 
become time dependent. 
To use these functions as a stable computational basis it is necessary to remove 
spin-spin interactions using special means \cite{Cory}. For example a WAHUHA multipulse 
sequence, consisting of four pulses, can be used for this purpose \cite{Waugh}.

\par In this paper it is shown how in the system of interacting spins with spin 
Z-components not commuting with total Hamiltonian, one can introduce a stable 
computational basis, which does not require the continuous application of a 
multipulse sequences. The trick is that instead of the correspondence ``one qubit'' 
- ``a pair of single spin states'' one makes the correspondence ``one qubit'' - 
``a pair of collective spin states''. Using the virtual spin formalism 
language \cite{Kes_Erm_Multiqubit_Spin_JETP_Lett} 
it can be written as ``one qubit'' - ``one virtual spin'', instead of ``one qubit'' - 
``one real spin''. Such approach is being developed in the framework of our program 
``Ouantum computer: many levels instead of many particles'' 
\cite{Kes_Erm_Virtual_Qubits_JETP}. Unlike  
the previous cases, in which the virtual spins have been defined on the levels 
of the individual multilevel particle - on the spins 3/2 or 7/2 levels 
\cite{Kes_Erm_Virtual_Qubits_JETP}-\cite{Kes_Erm_Physical_Implementation_JETP_Lett}, 
in the present case they are defined on the collective levels of the interacting 
spins system. It means that a qubit is ``delocalized'' and there is no direct 
correspondence ``one qubit'' - ``one particle''.  

\section{A simple model of two spins interaction}

\par A universal quantum gate set consists of single qubit rotations and 
two-qubit CNOT gate \cite{Barenco}. In order to implement such a set it will be enough 
to use a simple two spin interaction model. Let it will be the system of two non 
equivalent nuclear spins $I = 1/2$ and $S = 1/2$, connected by isotropic exchange 
interaction. 

\par This system Hamiltonian is

\begin{equation}\label{Hamiltonian}
\begin{array}{l}
{\cal H} = \hbar \omega _{0} (I_{z}+S_{z}) + \hbar \delta /2 (I_{z} - S_{z}) + \hbar J ({\bf I} \cdot {\bf S}),\\  
\omega _{0} = (1/2)(\gamma _I + \gamma _S) H_{0}, \delta = - (\gamma _I - \gamma _S) H_{0} 
\end{array}
\end{equation}

\noindent where $\gamma _I$ and $\gamma _S$ - the nuclei giromagnetic ratios, 
$J$ - the exchange integral, $H_{0}$ - the external constant magnetic field. 

\par This Hamiltonian eigenfunctions and corresponding eigenvalues are 

\begin{equation}\label{eigenfunctions}
\begin{array}{ll}
\ket{\psi _{1}} =  \ket{++} \equiv \ket{m_{I}= +1/2, m_{S}= +1/2}, 
& E_{1} \equiv \hbar \varepsilon _{1} = \hbar \omega _{0} + (1/4) \hbar J \\ 
\ket{\psi _{2}} = p \ket{+-} + q \ket{-+}, 
& E_{2} \equiv \hbar \varepsilon _{2} = - (1/4) \hbar J + (1/2) \hbar \theta \\  
\ket{\psi _{3}} = p \ket{-+} - q \ket{+-}, 
& E_{3} \equiv \hbar \varepsilon _{3} = - (1/4) \hbar J - (1/2) \hbar \theta \\  
\ket{\psi _{4}} = \ket{--}, 
& E_{4} \equiv \hbar \varepsilon _{4} = - \hbar \omega _{0} + (1/4) \hbar J   
\end{array}
\end{equation}

\noindent where  $p = cos( \phi /2), q = sin( \phi /2), \theta ^{2} = J^{2} + \delta ^{2}$,
$J/\delta = tg(\phi), 	-\pi/2 \leq \phi \leq \pi/2.$

\par This system has four allowed transitions with the following frequencies and 
relative intensities:

\begin{equation}\label{frequencies}
\begin{array}{ll}
\varepsilon _{12} = \omega _{0} + (1/2) J - (1/2) \theta,
& P_{12} \propto |\bra{\psi _{1}} I_{x}+S_{x} \ket{\psi _{2}} |^{2} = 1 + sin(\phi),\\
\varepsilon _{13} = \omega _{0} + (1/2) J + (1/2) \theta,
& P_{13} \propto |\bra{\psi _{1}} I_{x}+S_{x} \ket{\psi _{3}} |^{2} = 1 - sin(\phi),\\
\varepsilon _{24} = \omega _{0} - (1/2) J + (1/2) \theta,
& P_{24} \propto |\bra{\psi _{2}} I_{x}+S_{x} \ket{\psi _{4}} |^{2} = 1 + sin(\phi),\\
\varepsilon _{34} = \omega _{0} - (1/2) J - (1/2) \theta,
& P_{34} \propto |\bra{\psi _{3}} I_{x}+S_{x} \ket{\psi _{4}} |^{2} = 1 - sin(\phi),\\
\end{array}
\end{equation}

\par Fig. 1 depicts the energy levels of this system. This simple system allows exact 
solution and manifests the specific features and advantages of the proposed information 
coding -- the so-called ``virtual spin formalism''. In a case of the more general 
interaction (anisotropic exchange or dipole-dipole one) all necessary expressions can 
be obtained using perturbation theory. 

\section{Computational basis in the virtual spin representation}

\par The four dimensional Hilbert space, spanned on eigenfunctions (\ref{eigenfunctions}), 
can be considered as a direct product of the pair of virtual spins 
$R=1/2$ and $S=1/2$ two-dimensional Hilbert spaces \cite{Kes_Erm_Virtual_Qubits_JETP}. 
It means that the 
eigenfunctions (\ref{eigenfunctions}) of Hamiltonian (\ref{Hamiltonian}) can be taken 
to form  the computational  basis for the two qubit system, thus:

\begin{equation}\label{comp_basis}
\begin{array}{ll}
\ket{00} = \ket{\psi _{1}}, &\ket{01} = \ket{\psi _{2}},\\ 
\ket{10} = \ket{\psi _{3}}, &\ket{11} = \ket{\psi _{4}}.  
\end{array}
\end{equation}

\par This notation means that the qubit corresponds to the virtual spin, so: 

\begin{equation}\label{corespondance}
\begin{array}{ll}
\ket{0_{Q}} = \ket{m_{Q}= -1/2}, & \ket{1_{Q}} = \ket{m_{Q}= +1/2},\\
\ket{0_{R}} = \ket{m_{R}= -1/2}, & \ket{1_{R}} = \ket{m_{R}= +1/2}.   
\end{array}
\end{equation}

\noindent and two qubit states correspond to the computational basis:

\begin{equation}\label{comp_basis_2}
\begin{array}{ll}
\ket{00} = \ket{m_{Q}= -1/2,m_{R}= -1/2},&\ket{01} = \ket{m_{Q}= -1/2,m_{R}= +1/2},\\
\ket{10} = \ket{m_{Q}= +1/2,m_{R}= -1/2},&\ket{11} = \ket{m_{Q}= +1/2,m_{R}= +1/2}.
\end{array}
\end{equation}

\par Now a resonance transition between any two real states of physical 
system admits the interpretation as virtual spin reorientations. For example, 
the irradiation of the transition $\bra{\psi _{1}}\leftrightarrow \ket{\psi _{2}}$ 
can be interpreted in the virtual 
spin representation as the Q spin rotation etc. Using the previous 
results \cite{Kes_Erm_Multiqubit_Spin_JETP_Lett}-\cite{Kes_Erm_Virtual_Qubits_JETP} 
it can be shown, that in this two qubit system the universal 
gates set can be realized by means of the following resonance electromagnetic
field pulses.

\begin{center}
\begin{tabular}{|c|l|}
\hline
\ {\bf Logic operation:} &  {\bf Realization by transition(s):}\\  
Virtual spin Q rotation 
& $\bra{\psi _{1}}\leftrightarrow \ket{\psi _{2}}$ and $\bra{\psi _{3}}\leftrightarrow \ket{\psi _{4}}$\\    
Virtual spin R rotation 
& $\bra{\psi _{1}}\leftrightarrow \ket{\psi _{3}}$ and $\bra{\psi _{2}}\leftrightarrow \ket{\psi _{4}}$\\    
Controlled $Q$ spin inversion $CNOT_{R \rightarrow Q}$ 
& $\pi$ - pulse applied to $\bra{\psi _{3}}\leftrightarrow \ket{\psi _{4}}$\\
Controlled $R$ spin inversion $CNOT_{Q \rightarrow R}$
& $\pi$ - pulse applied to  $\bra{\psi _{2}}\leftrightarrow \ket{\psi _{4}}$
\cr \hline
\end{tabular}
\end{center}

\noindent where $CNOT_{R \rightarrow Q}$ means that qubit $Q$ undergoes the $NOT$ operation, 
which is controlled by the state of qubit $R$, and $\pi$ - pulse means the virtual 
spin $Q$ rotation through the angle $\pi$. The corresponding resonance pulses are 
depicted on Fig. 2. It can be seen that all quantum gates can be implemented 
using just one pulse: for virtual spin rotation a double frequency pulse is 
necessary, whereas for $CNOT$ operation - a single frequency pulse. 

\section{Conclusions}

\par It was shown that the information coding onto virtual spins allows to implement 
a universal gate set in a solid state two interacting spin system. The advantages of 
this coding are connected with the fact, that spin-spin interactions can be large 
and that there is no need to use continuous irradiation with a multipulse sequence 
to have a stable computational basis. Large spin-spin interaction produces big 
resonance frequencies differences, which facilitate the selective resonance 
excitation of the individual transitions, desired for gates implementation. In 
addition, the gate operation time is under full control of an experimentalist and 
can be done short, whereas the coding using real spins requires time, defined by 
exchange interaction value (which is a given molecule property) and can be rather 
long. 

\par It should be noted that the suggested approach can be useful also for 
LS NMR QIP in a case, when an exchange interaction is not averaged to spin 
Z-components. It can be used also for information coding onto any cluster 
of interacting particles of arbitrary nature. The only requirements are the 
existence of the proper selection rules for resonance transitions among 
cluster stationary states, on which a two virtual qubit system is defined. 
An example of the virtual qubit formalism applied to optical states of a 
single atom is given in the paper \cite{Erm_Kes_Sam_Four_atomic_optical_SPIE}.

\end{document}